# A design of magnetic tunnel junctions for the deployment of neuromorphic hardware for edge computing

Davi Rodrigues[1,*], Eleonora Raimondo[2,3], Riccardo Tomasello[1], Mario Carpentieri[1], Giovanni Finocchio[3,*]

[1] *Department of Electrical and Information Engineering, Politecnico di Bari, 70126, Bari, Italy*

[2] *Istituto Nazionale di Geofisica e Vulcanologia, Via di Vigna Murata 605, 00143 Rome, Italy*

[3] *Department of Mathematical and Computer Sciences, Physical Sciences and Earth Sciences, University of Messina, 98166, Messina, Italy*

**Abstract**

The electrically readable complex dynamics of robust and scalable magnetic tunnel junctions (MTJs) offer promising opportunities for advancing neuromorphic computing. In this work, we present an MTJ design with a free layer and two polarizers capable of computing the sigmoidal activation function and its gradient at the device level. This design enables both feedforward and backpropagation computations within a single device, extending neuromorphic computing frameworks previously explored in the literature by introducing the ability to perform backpropagation directly in hardware. Our algorithm implementation reveals two key findings: (i) the small discrepancies between the MTJ-generated curves and the exact software-generated curves have a negligible impact on the performance of the backpropagation algorithm, (ii) the device implementation is highly robust to inter-device variation and noise, and (iii) the proposed method effectively supports transfer learning and knowledge distillation. To demonstrate this, we evaluated the performance of an edge computing network using weights from a software-trained model implemented with our MTJ design. The results show a minimal loss of accuracy of only 0.1% for the Fashion MNIST dataset and 2% for the CIFAR-100 dataset compared to the original software implementation. These results highlight the potential of our MTJ design for compact, hardware-based neural networks in edge computing applications, particularly for transfer learning.

Corresponding authors: *gfinocchio@unime.it, *davi.rodrigues@poliba.it

**Introduction**

Edge devices are increasingly integrated into various aspects of society, from smart homes and autonomous vehicles to surveillance cameras and intelligent manufacturing robots (1–3). Edge computing offers several advantages over centralized computing, including low latency, enhanced privacy, bandwidth efficiency, and scalability. However, the high accuracy of deep learning models comes at the cost of significant computational and storage requirements, making implementation on edge devices challenging. Training deep learning models is computationally intensive and space consuming due to the millions of parameters that must be iteratively refined over numerous epochs (4, 5). These challenges are compounded by the limitations of current technology, which struggles to address the nonlinearity, high connectivity, and memory requirements of deep learning, posing significant obstacles to traditional Von Neumann computing architectures.

A fundamentally resource-intensive task is backpropagation (6–8). It requires the backward propagation of information through the network and the storage of parameters, outputs, and gradients for each instance. Although many alternative algorithms have been proposed, including those based on optimal control theory (9, 10), physics-inspired concepts (11, 12), local learning (13, 14), and reservoir computing (15, 16) - all of which theoretically bypass complex training algorithms - the backpropagation algorithm remains the most widely used. This is due to its proven versatility, robustness, and effectiveness despite the challenges it presents.

Specialised hardware solutions can integrate in-situ memory and introduce non-linear behaviour to improve the computational efficiency of backpropagation algorithms. This concept has shown promising results in photonic implementations (17–19). However, photonic implementations face significant challenges due to the need for fully dedicated hardware and scalability issues. Spintronic devices, known for their CMOS compatibility, non-volatility, non-linearity, robustness, and tunability, have emerged as promising candidates for seamless integration into hardware implementations of neural networks, effectively performing various functions (15, 20, 21). A fundamental building block of this technology is the magnetic tunnel junction (MTJ), which, through the magnetoresistance effect, converts the relative orientation of two magnetisation vectors into an electrical resistance value (22, 23). MTJs are scalable, robust nanodevices with operation voltages in the range of millivolts and have shown great success to emulate memristor behaviour as well as to perform conventional activation functions (AFs) and spikes in spiking neural networks (21, 24–27).

In this work, we exploit both the non-linear response of MTJs to develop and present a technique for on-site efficiently computing non-linear AFs and their gradients for backpropagation algorithms in a highly parallel manner. Gradient computations are also crucial for other training algorithms such as direct feedback alignment (13), adjoint methods (28), stochastic gradient descent (29), and natural gradient descent (30). The proposed device integrates seamlessly with the hardware implementation of artificial neural networks (ANNs), allowing simultaneous computation and storage of the necessary parameters for both forward and backward passes. We used micromagnetic simulations to numerically calculate the device behaviour using experimentally obtained data (31–34). Notably, we found no significant loss of accuracy, always less than 3%, in networks fully implemented with the MTJ devices and trained using backpropagation. This highlights that quasi-exact gradient computation allows the ANN to maintain high levels of accuracy while significantly reducing memory and processing costs. The device implementation also demonstrated significant resilience to noise and device-to-device variation. The proposed device is well suited for transfer learning and edge computing. It enables the use of knowledge from larger, deeper networks trained in software for on-chip realisation of compact ANNs (35–37).

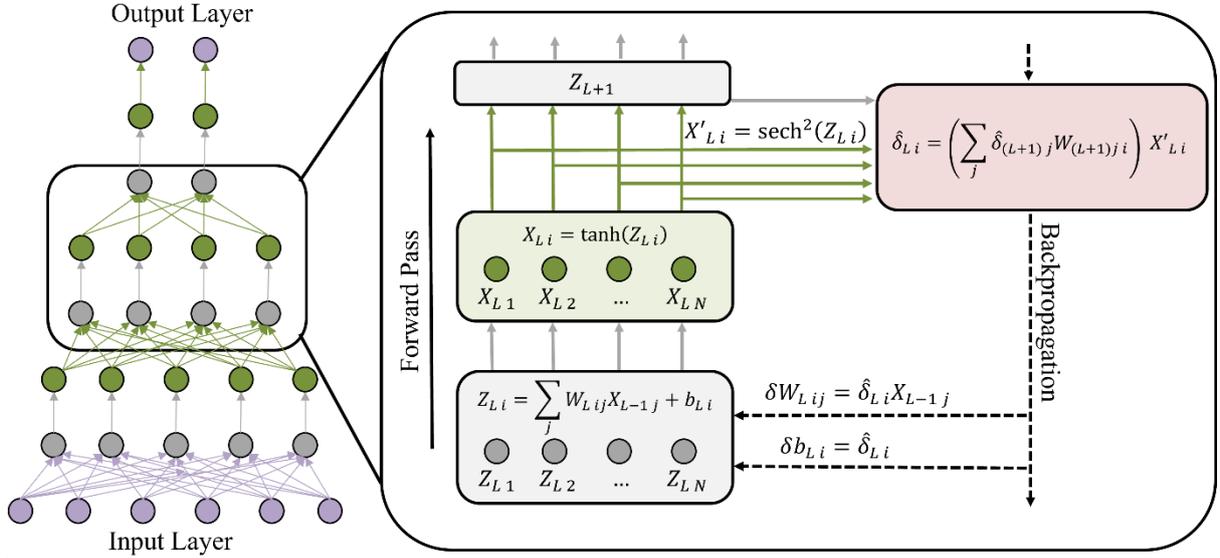

**FIG. 1. Backpropagation algorithm.** A schematic of backpropagation implementation of a simple ANN. Gray nodes, labeled $Z_{Ln}$ where $L$ and $n$ are layer and neuron, respectively, represent neurons that execute linear functions. Green nodes, labeled $X_{Ln}$ represent neurons in the non-linear layers. During the forward pass, information flows sequentially from the input layer to the output layer, as shown in the inset on the right. During the backward pass, parameters within the linear layers are adjusted based on the gradient of the loss function. To provide a concrete example, we explicitly show the update mechanism for the linear parameters within a layer, using the cumulative information from previous layers, as shown in the pink box. The update function requires the output values, AF derivatives, and previous linear parameters.

**Backpropagation algorithm**

The backpropagation algorithm is a multi-step process used to iteratively update the parameters of an ANN, consisting of a forward pass, a loss computation, and a backward pass (7). During the forward pass, data is propagated sequentially through each layer of the network, see Fig. 1. At each layer, neurons either compute linear combinations of outputs from previous layers or apply nonlinear AFs. In this way, the output layer generates a set of values, associated to the prediction of the ANN, based on the set input values. A loss function is then computed to quantify the performance of the network, measuring its ability to correctly classify inputs.

In the backward pass, the parameters are updated to minimize the loss function and improve accuracy. This is achieved by adjusting the linear weights according to the gradient of the loss function with respect to each parameter, following the rule of gradient descent. The derivative calculations propagate backward through the network via the chain rule, meaning that the update for each layer is influenced by the updates in subsequent layers, see the inset in Fig. 1. This algorithm allows for accurate and deterministic updates, effectively reducing the loss function.

However, backpropagation poses a significant memory challenge, as it requires the storage of intermediate values and their corresponding gradients for each node in the network. A promising solution to this challenge is the hybrid CMOS-MTJ implementation, which enables in-situ storage of these values within the nodes, potentially reducing memory requirements and improving efficiency.

**Implementation of the backpropagation algorithm with MTJs.**

Previous research has demonstrated the feasibility of encoding the hyperbolic tangent (tanh) function along a component of the magnetization in a MTJ (26, 38, 39). It exploits the tunnelling magnetoresistance effect, where the resistance across the MTJ depends on the projection of the magnetization direction of the free layer (FL) onto that of the reference layer (RL). This mechanism allows different components of the FL magnetization to be converted into an electrical signal. By manipulating the direction of the FL magnetization via magnetic fields, electric currents, strain, or voltage-controlled parameters (22–24, 40–45), we can obtain a tanh response of the resistance through the device as a function of external control parameters (26, 40).

A main result of this manuscript is that, using a simple identity, when the magnetization dynamics are confined within a plane - achievable by appropriate anisotropy design - the magnetization in the perpendicular direction encodes the hyperbolic secant (sech) function. The latter corresponds to the square of the gradient function of the tanh function. In simpler terms, if $\mathbf{m_x}(I) = \tanh(I)$, where $I$ is the applied current on the field line, then $\mathbf{m_z}(I) \approx \sqrt{1 - \mathbf{m_x}^2(I)} = \text{sech}(I)$. And, since $(\tanh(I))' = \text{sech}^2(I)$, this implies that $\mathbf{m_z}(I)^2 \approx (\tanh(I))'$. This allows to obtain both the tanh AF and its gradient by manipulating the FL magnetization, see Fig. 2(a)-(c).

A single MTJ can provide both the AF and its associated gradient from by measuring two orthogonal FL magnetization components. This can be achieved by considering a three terminal device presenting two RL with different magnetization directions, allowing the measurement of different FL components (46). Another method is to change the orientation of the polarizer using magneto-ionics.(47, 48) With this approach the polarizer can be set along the in-plane direction when the ANN work in forward pass and along the out-of-plane configuration for backward pass.

Fig. 2(b)-(c) compare the micromagnetic simulation results for the magnetization components in the device proposed in Fig. 2(a) with the ideal curves. For details about the micromagnetic simulations, see Supplemental Material A. The curve representing the *x*-component of the magnetization closely resembles the tanh function, while the curve for the *z*-component tends to the square root of the gradient. This behaviour underscores the suitability of the proposed device implementation as nonlinear nodes in ANNs, promising significant reductions in memory and processing costs.

We emphasize that the functional response of MTJs can be finely tuned by careful engineering of material properties and geometry, providing significant flexibility to achieve optimal behaviour. In our study, we observed sharp changes in the resistance curve near saturation points, which could potentially lead to vanishing gradient problems during the learning process. However, the adaptability of the ANN, combined with appropriate network design and backpropagation updates, proved effective in mitigating these variations. To demonstrate the capabilities of the proposed device, we conducted a comprehensive study involving its integration into different configurations and topologies of ANNs.

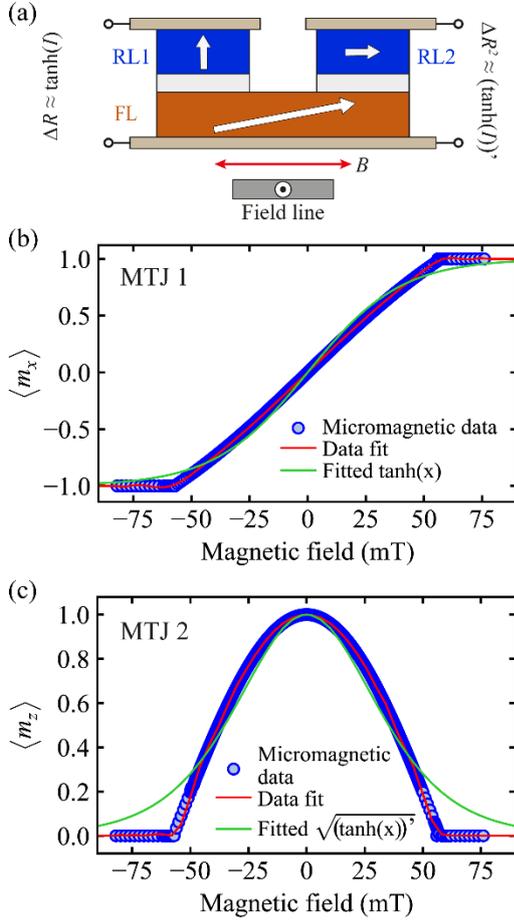

**FIG. 2. Implementation of MTJ-based simultaneous calculation of the activation function and gradient.** (a) Schematic of the proposed device featuring a single FL and two RLs. RL 1 measures the component along *x*, while RL 2 measures the component along *z*. An external magnetic field generated by an electric current sets the overall magnetization direction in the FL. (b) Functional response measured by RL 1, associated to the tanh AF. (c) Functional response measured by RL 2, associated to the gradient of the AF. In (b) and (c), blue dots represent micromagnetic simulation results, red lines show numerical piecewise polynomial fits for continuous estimation, and green lines show fits with the ideal AF and its gradient. Details of the micromagnetic simulations can be found in the supplementary material.

**Results and Discussion**

Fig. 3(a) shows the schematic of the proposed network, where the nonlinear nodes are implemented using the device shown in Fig. 2(a). This implementation allows parallel execution of both forward and backward passes without relying on external memory, which is essential for compact edge applications. We show that deviations from exact gradients in hardware have minimal impact on the performance of the additional on-chip training of the neural network. This is the second main result of this manuscript, as it indicates that while the use of non-exact gradients significantly improves network efficiency, it does not compromise accuracy. We have evaluated the performance of our neuron implementation using two distinct ANN architectures for image classification tasks to assess the device's performance and scalability. We emphasize that our focus in this work is not on achieving peak performance in accuracy. Instead, we have aimed for comparable accuracy while ensuring the use of accurate and quasi-accurate gradients. Specifically, we conducted experiments on the Fashion MNIST and CIFAR-100 datasets. For the Fashion MNIST dataset, we employed a simple convolutional neural network (CNN) consisting of two convolutional blocks, each comprising a convolutional layer followed by an AF layer and a max-pooling layer. Subsequently, a straightforward feedforward layer was applied, comprising a linear layer followed by an AF layer and another linear layer. In total, the CNN had 421642 parameters. In contrast, for the CIFAR-100 dataset, we utilized a DenseNet architecture (49). This DenseNet configuration comprised three dense blocks and two transition blocks. The growth rate was set to 12, and the block configuration was defined as (16, 16, 16). The total number of parameters for this DenseNet architecture reached 561052. For all networks, we used cross-entropy as the loss function and the Adam optimization algorithm (50) with a learning rate of 0.001.

In our evaluation of both networks, we compared ideal scenarios using exact AFs and gradients with scenarios using the AF and quasi-exact gradient generated by the proposed device. Various combinations were explored to evaluate the accuracy of the MTJ-implemented neural network against state-of-the-art models. Fig. 3(b)-(c) show the evolution of the cost function and accuracy during training for the different networks and tasks. For smaller networks, such as the one used to classify Fashion MNIST, there was no significant difference between using exact and quasi-exact functions. However, for larger networks, such as the one used for CIFAR-100, the difference was more noticeable, although high accuracy was still achieved.

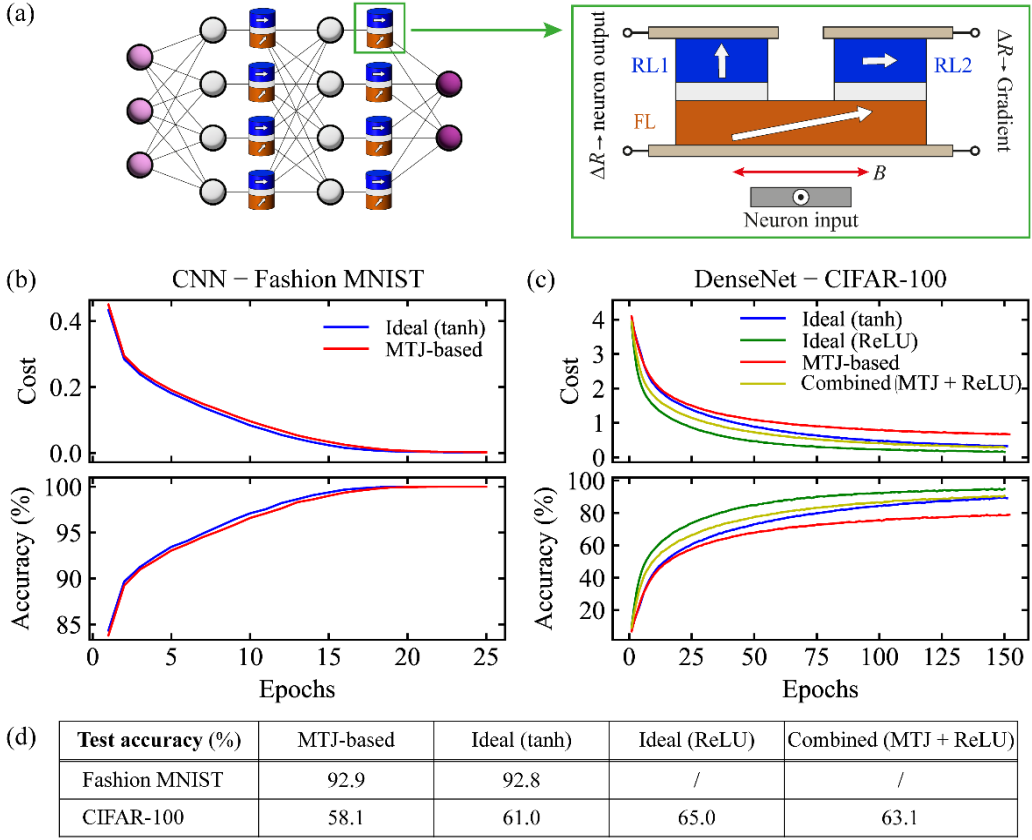

FIG. 3. **A comparison between ideal ANNs and MTJ-based ANNs.** (a) Schematic of the envisioned on-chip MTJ-based neural network application, where each nonlinear node corresponds to a device capable of reading the two magnetization components in the FL in the MTJs. The input is provided by the current along the field line, and the resistances along each component provide the AF and associated gradient. (b) Comparison of cost and accuracy during training for a CNN classifying Fashion MNIST, considering both the ideal network and the MTJ-based network. (c) Comparison of cost and accuracy for the DenseNet network classifying CIFAR-100. Four different networks are considered: one with ReLU AFs and exact gradients, one with tanh AFs and exact gradients, one simulation using the MTJ-based AF for all AFs and gradients, and a hybrid configuration with MTJ-based AFs in the first two dense blocks and ReLU in the last dense block. (d) Table comparing the accuracies obtained during testing.

Combined approaches were able to approximate the ideal exact case while potentially significantly reducing power and memory costs. The lower efficiency observed in larger networks with MTJ-based gradients may be related to the vanishing gradient problem, where the gradient decreases rapidly at the tails. Proper network topology design, such as dense blocks and implementation of dropout techniques, can help mitigate this problem (49, 51, 52).

The table in Fig. 3(d) shows the accuracy obtained during testing, indicating no significant drop compared to ideal networks trained with exact gradients in software. For the Fashion MNIST, the accuracy drop was only about 0.1% while for CIFAR-100, the combined design resulted in a drop of just about 2%. Although the MTJ-based network was trained using conventional techniques, we believe that hardware-aware training methods could further reduce the accuracy gap between ideal and MTJ-implemented ANNs.

The third key result is the demonstrated resilience of the proposed hardware implementation to noise and device-to-device variation. We ran simulations with Gaussian noise applied independently to each node and instance, varying the noise amplitude up to 100% of the maximum magnetization component value. Fig. 4(a)-(b) illustrate the evolution of the cost function and accuracy during training for the Fashion MNIST classification task, while Figure 4(c) summarizes the test accuracy. Notably, the accuracy did not degrade for noise levels up to 10%, and in some cases the MTJ-based network even outperformed the ideal scenario, highlighting its robustness.

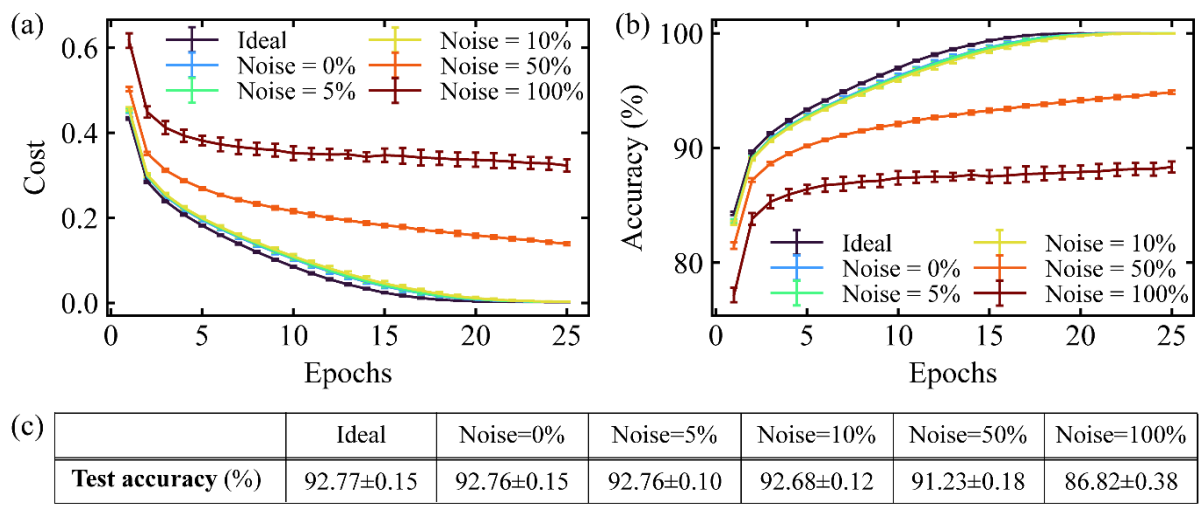

|  | Ideal | Noise=0% | Noise=5% | Noise=10% | Noise=50% | Noise=100% |
|---|---|---|---|---|---|---|
| **Test accuracy** (%) | 92.77±0.15 | 92.76±0.15 | 92.76±0.10 | 92.68±0.12 | 91.23±0.18 | 86.82±0.38 |

**FIG. 4. Performance resilience of the device implementation.** (a)-(b) Comparison of cost and accuracy for Fashion MNIST classification under different noise levels. Noise is expressed as the ratio of the maximum value of the magnetization component. Each noise level was tested with 10 simulations, all starting from the same initial configuration. (c) Summary table comparing the accuracies obtained during the tests.

Given the demonstrated effectiveness of MTJ-based ANNs in small network architectures, we explored their application in transfer learning based on knowledge distillation (KD), a process where the knowledge contained in a large, complex model (the teacher model) is transferred to a smaller and more compact model (the student model) (53, 54). This transfer of knowledge is performed by adjusting the loss of the student model to incorporate predictions from the teacher model (54, 55). For details, see Supplemental Material B. In our evaluation, we considered the task of classifying the CIFAR-10 dataset (56). The teacher model chosen was ResNet-18 (57), with which we achieved an accuracy of 82%. For the student model, we chose a simple CNN consisting of three convolutional blocks (each consisting of a convolutional layer, a non-linear layer and a max-pooling layer) and a fully connected layer. We examined both the ReLU AF and the MTJ-based AF and gradient. During the knowledge distillation process, the training sessions for the different networks started with identical parameter sets. Fig. 5(a)-(b) shows a comparison of the cost and accuracy achieved by the student model with and without KD. It's worth noting that the loss for KD includes the soft loss associated with the teacher model. Notably, the MTJ-based network consistently shows improved accuracy compared to the scenario without KD. Furthermore, it tends to approach the accuracy of the ideal network with accurate AF and gradient. These results highlight the effectiveness of MTJ-based compact networks in transfer learning contexts.

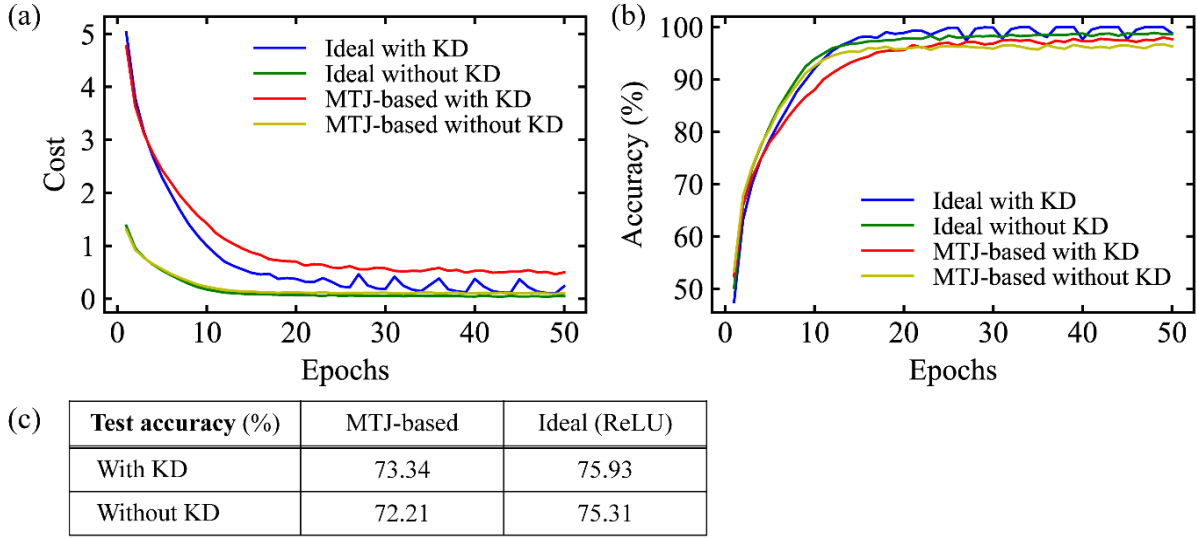

FIG. 5. **Evaluating knowledge distillation for MTJ-based ANNs.** (a) Cost during training is compared with and without KD for both the ideal network and the MTJ-based network. (b) Accuracy during training and testing is compared with and without KD for both the ideal network and the MTJ-based network.

**V. Conclusion and Outlook**

In this manuscript, we propose an MTJ-based hardware solution for efficient and accurate feedforward and backward computation in compact edge applications. By exploiting the nonlinear response of MTJs - known for their robustness, compactness, and low power consumption - we achieve simultaneous computation of AFs and gradients.

Our evaluation, conducted on classification tasks using the Fashion MNIST and CIFAR-100 datasets, demonstrates three key results: the small discrepancies between MTJ-generated and exact software-generated curves have a negligible impact on the performance of the backpropagation algorithm, and the hardware implementation is highly robust to noise and inter-device variation. We further validate the versatility of our implementation through transfer learning and knowledge distillation experiments on the CIFAR-10 dataset, showing that MTJ-based networks achieve accuracies comparable to those of ideally accurate models. This highlights the potential of our approach for optimizing edge computing, where compactness and efficiency are paramount.

Overall, our results, combined with ongoing advances in MTJs and spintronic technologies, promise to significantly improve the performance of ANNs for edge applications, addressing key challenges that have limited their widespread adoption.

**Acknowledgements**

This work was supported by the project number 101070287 - SWAN-on-chip - HORIZON-CL4-2021-DIGITAL-EMERGING-01, the project PRIN 2020LWPKH7 "The Italian factory of micromagnetic modelling and spintronics" and the project PRIN20222N9A73 "SKYrmion-based magnetic tunnel junction to design a temperature SENSor-SkySens", funded by the Italian Ministry of University and Research (MUR) and by the PETASPIN Association (www.petaspin.com). DR, RT and MC acknowledge the support from the project PE0000021, "Network 4 Energy Sustainable Transition - NEST", funded by the European Union - NextGenerationEU, under the National Recovery and Resilience Plan (NRRP), Mission 4 Component 2 Investment 1.3 - Call for Tender No. 1561 dated 11.10.2022 of the Italian MUR (CUP C93C22005230007). DR also acknowledges the support of the project D.M. 10/08/2021 n. 1062 (PON Ricerca e Innovazione), funded by the Italian MUR, and ER acknowledges the support of the project PON Capitale Umano (CIR_00030), funded by the Italian MUR.

**References**

1. K. Cao, Y. Liu, G. Meng, Q. Sun, An Overview on Edge Computing Research. *IEEE Access* **8**, 85714–85728 (2020).

2. M. Satyanarayanan, The Emergence of Edge Computing. *Computer (Long Beach Calif)* **50**, 30–39 (2017).

3. W. Shi, J. Cao, Q. Zhang, Y. Li, L. Xu, Edge Computing: Vision and Challenges. *IEEE Internet Things J* **3**, 637–646 (2016).

4. H. Larochelle, Y. Bengio, J. Louradour, L. U. Ca, Exploring Strategies for Training Deep Neural Networks. *The Journal of Machine Learning Research* **1**, 1–40 (2009).

5. I. N. da Silva, D. Hernane Spatti, R. Andrade Flauzino, L. H. B. Liboni, S. F. dos Reis Alves, "Artificial Neural Network Architectures and Training Processes" in *Artificial Neural Networks* (Springer International Publishing, Cham, 2017; http://link.springer.com/10.1007/978-3-319-43162-8_2), pp. 21–28.

6. J. Chen, X. Ran, Deep Learning With Edge Computing: A Review. *Proceedings of the IEEE* **107**, 1655–1674 (2019).

7. Y. LeCun, Y. Bengio, G. Hinton, Deep learning. *Nature* **521**, 436–444 (2015).

8. I. H. Sarker, Machine Learning: Algorithms, Real-World Applications and Research Directions. *SN Comput Sci* **2**, 160 (2021).

9. D. R. Rodrigues, E. Raimondo, V. Puliafito, R. Moukhadder, B. Azzerboni, A. Hamadeh, P. Pirro, M. Carpentieri, G. Finocchio, Dynamical Neural Network Based on Spin Transfer Nano-Oscillators. *IEEE Trans Nanotechnol* **22**, 800–805 (2023).

10. G. Furuhata, T. Niiyama, S. Sunada, Physical Deep Learning Based on Optimal Control of Dynamical Systems. *Phys Rev Appl* **15**, 034092 (2021).

11. L. G. Wright, T. Onodera, M. M. Stein, T. Wang, D. T. Schachter, Z. Hu, P. L. McMahon, Deep physical neural networks trained with backpropagation. *Nature* **601**, 549–555 (2022).

12. M. Hermans, M. Burm, T. Van Vaerenbergh, J. Dambre, P. Bienstman, Trainable hardware for dynamical computing using error backpropagation through physical media. *Nat Commun* **6**, 6729 (2015).

13. A. Nøkland, Direct Feedback Alignment Provides Learning in Deep Neural Networks. arXiv:1609.01596 (2016).

14. L. Bottou, V. Vapnik, Local Learning Algorithms. *Neural Comput* **4**, 888–900 (1992).

15. G. Finocchio, J. A. C. Incorvia, J. S. Friedman, Q. Yang, A. Giordano, J. Grollier, H. Yang, F. Ciubotaru, A. V Chumak, A. J. Naeemi, S. D. Cotofana, R. Tomasello, C. Panagopoulos,

M. Carpentieri, P. Lin, G. Pan, J. J. Yang, A. Todri-Sanial, G. Boschetto, K. Makasheva, V. K. Sangwan, A. R. Trivedi, M. C. Hersam, K. Y. Camsari, P. L. McMahon, S. Datta, B. Koiller, G. H. Aguilar, G. P. Temporão, D. R. Rodrigues, S. Sunada, K. Everschor-Sitte, K. Tatsumura, H. Goto, V. Puliafito, J. Åkerman, H. Takesue, M. Di Ventra, Y. V Pershin, S. Mukhopadhyay, K. Roy, I.- Ting Wang, W. Kang, Y. Zhu, B. K. Kaushik, J. Hasler, S. Ganguly, A. W. Ghosh, W. Levy, V. Roychowdhury, S. Bandyopadhyay, Roadmap for unconventional computing with nanotechnology. *Nano Futures* **8**, 012001 (2024).

16. G. Tanaka, T. Yamane, J. B. Héroux, R. Nakane, N. Kanazawa, S. Takeda, H. Numata, D. Nakano, A. Hirose, Recent advances in physical reservoir computing: A review. *Neural Networks* **115**, 100–123 (2019).

17. S. Pai, Z. Sun, T. W. Hughes, T. Park, B. Bartlett, I. A. D. Williamson, M. Minkov, M. Milanizadeh, N. Abebe, F. Morichetti, A. Melloni, S. Fan, O. Solgaard, D. A. B. Miller, Experimentally realized in situ backpropagation for deep learning in photonic neural networks. *Science (1979)* **380**, 398–404 (2023).

18. T. W. Hughes, M. Minkov, Y. Shi, S. Fan, Training of photonic neural networks through in situ backpropagation and gradient measurement. *Optica* **5**, 864 (2018).

19. X. Guo, T. D. Barrett, Z. M. Wang, A. I. Lvovsky, Backpropagation through nonlinear units for the all-optical training of neural networks. *Photonics Res* **9**, B71 (2021).

20. J. Grollier, D. Querlioz, K. Y. Camsari, K. Everschor-Sitte, S. Fukami, M. D. Stiles, Neuromorphic spintronics. *Nat Electron* **3**, 360–370 (2020).

21. G. Finocchio, M. Di Ventra, K. Y. Camsari, K. Everschor-Sitte, P. Khalili Amiri, Z. Zeng, The promise of spintronics for unconventional computing. *J Magn Magn Mater* **521**, 167506 (2021).

22. N. Maciel, E. Marques, L. Naviner, Y. Zhou, H. Cai, Magnetic Tunnel Junction Applications. *Sensors* **20**, 121 (2019).

23. J.-G. (Jimmy) Zhu, C. Park, Magnetic tunnel junctions. *Materials Today* **9**, 36–45 (2006).

24. D. R. Rodrigues, R. Moukhader, Y. Luo, B. Fang, A. Pontlevy, A. Hamadeh, Z. Zeng, M. Carpentieri, G. Finocchio, Spintronic Hodgkin-Huxley-Analogue Neuron Implemented with a Single Magnetic Tunnel Junction. *Phys Rev Appl* **19**, 064010 (2023).

25. J. Torrejon, M. Riou, F. A. Araujo, S. Tsunegi, G. Khalsa, D. Querlioz, P. Bortolotti, V. Cros, K. Yakushiji, A. Fukushima, H. Kubota, S. Yuasa, M. D. Stiles, J. Grollier, Neuromorphic computing with nanoscale spintronic oscillators. *Nature* **547**, 428–431 (2017).

26. E. Raimondo, A. Giordano, A. Grimaldi, V. Puliafito, M. Carpentieri, Z. Zeng, R. Tomasello, G. Finocchio, Reliability of Neural Networks Based on Spintronic Neurons. *IEEE Magn Lett* **12**, 10–14 (2021).

27. N. Locatelli, A. F. Vincent, A. Mizrahi, J. S. Friedman, D. Vodenicarevic, J.-V. Kim, J.-O. Klein, W. Zhao, J. Grollier, D. Querlioz, "Spintronic Devices as Key Elements for Energy-Efficient Neuroinspired Architectures" in *Design, Automation & Test in Europe Conference & Exhibition (DATE), 2015* (IEEE Conference Publications, New Jersey, 2015; http://ieeexplore.ieee.org/xpl/articleDetails.jsp?arnumber=7092535)vols. 2015-April, pp. 994–999.

28. L. S. Pontryagin, *Mathematical Theory of Optimal Processes* (Routledge, 2018).

29. L. Bottou, "Large-Scale Machine Learning with Stochastic Gradient Descent" in *Proceedings of COMPSTAT'2010* (Physica-Verlag HD, Heidelberg, 2010), pp. 177–186.

30. J. Martens, D. London, "New Insights and Perspectives on the Natural Gradient Method" (2020); https://doi.org/10.5555/3666122.3668471.

31. L. Lopez-Diaz, D. Aurelio, L. Torres, E. Martinez, M. A. Hernandez-Lopez, J. Gomez, O. Alejos, M. Carpentieri, G. Finocchio, G. Consolo, Micromagnetic simulations using Graphics Processing Units. *J Phys D Appl Phys* **45**, 323001 (2012).

32. A. Giordano, G. Finocchio, L. Torres, M. Carpentieri, B. Azzerboni, Semi-implicit integration scheme for Landau–Lifshitz–Gilbert-Slonczewski equation. *J Appl Phys* **111**, 07D112 (2012).

33. P. Khalili Amiri, Z. M. Zeng, J. Langer, H. Zhao, G. Rowlands, Y. J. Chen, I. N. Krivorotov, J. P. Wang, H. W. Jiang, J. A. Katine, Y. Huai, K. Galatsis, K. L. Wang, Switching current reduction using perpendicular anisotropy in CoFeB-MgO magnetic tunnel junctions. *Appl Phys Lett* **98** (2011).

34. C. Chappert, A. Fert, F. N. Van Dau, The emergence of spin electronics in data storage. *Nanoscience and Technology: A Collection of Reviews from Nature Journals*, 147–157 (2009).

35. W. Shi, J. Cao, Q. Zhang, Y. Li, L. Xu, Edge Computing: Vision and Challenges. *IEEE Internet Things J* **3**, 637–646 (2016).

36. Y. Mao, C. You, J. Zhang, K. Huang, K. B. Letaief, A Survey on Mobile Edge Computing: The Communication Perspective. *IEEE Communications Surveys & Tutorials* **19**, 2322–2358 (2017).

37. K. Cao, Y. Liu, G. Meng, Q. Sun, An Overview on Edge Computing Research. *IEEE Access* **8**, 85714–85728 (2020).

38. S. S. P. Parkin, C. Kaiser, A. Panchula, P. M. Rice, B. Hughes, M. Samant, S.-H. Yang, Giant tunnelling magnetoresistance at room temperature with MgO (100) tunnel barriers. *Nat Mater* **3**, 862–867 (2004).

39. L. E. Nistor, B. Rodmacq, C. Ducruet, C. Portemont, I. L. Prejbeanu, B. Dieny, Correlation Between Perpendicular Anisotropy and Magnetoresistance in Magnetic Tunnel Junctions. *IEEE Trans Magn* **46**, 1412–1415 (2010).

40. J. Cai, B. Fang, L. Zhang, W. Lv, B. Zhang, T. Zhou, G. Finocchio, Z. Zeng, Voltage-Controlled Spintronic Stochastic Neuron Based on a Magnetic Tunnel Junction. *Phys Rev Appl* **11**, 034015 (2019).

41. G. D. Fuchs, J. A. Katine, S. I. Kiselev, D. Mauri, K. S. Wooley, D. C. Ralph, R. A. Buhrman, Spin torque, tunnel-current spin polarization, and magnetoresistance in MgO magnetic tunnel junctions. *Phys Rev Lett* **96**, 1–4 (2006).

42. S. Kanai, F. Matsukura, H. Ohno, Electric-field-induced magnetization switching in CoFeB/MgO magnetic tunnel junctions with high junction resistance. *Appl Phys Lett* **108**, 192406 (2016).

43. S. Ikeda, J. Hayakawa, Y. M. Lee, F. Matsukura, Y. Ohno, T. Hanyu, H. Ohno, Magnetic tunnel junctions for spintronic memories and beyond. *IEEE Trans Electron Devices* **54**, 991–1002 (2007).

44. A. A. Tulapurkar, Y. Suzuki, A. Fukushima, H. Kubota, H. Maehara, K. Tsunekawa, D. D. Djayaprawira, N. Watanabe, S. Yuasa, Spin-torque diode effect in magnetic tunnel junctions. *Nature* **438**, 339–342 (2005).

45. T. Nozaki, Y. Shiota, M. Shiraishi, T. Shinjo, Y. Suzuki, Voltage-induced perpendicular magnetic anisotropy change in magnetic tunnel junctions. *Appl Phys Lett* **96**, 15–18 (2010).

46. P. M. Braganca, J. A. Katine, N. C. Emley, D. Mauri, J. R. Childress, P. M. Rice, E. Delenia, D. C. Ralph, R. A. Buhrman, A Three-Terminal Approach to Developing Spin-Torque Written Magnetic Random Access Memory Cells. *IEEE Trans Nanotechnol* **8**, 190–195 (2009).

47. A. E. Kossak, D. Wolf, G. S. D. Beach, Magneto-ionic enhancement and control of perpendicular magnetic anisotropy. *Appl Phys Lett* **121** (2022).

48. U. Bauer, L. Yao, A. J. Tan, P. Agrawal, S. Emori, H. L. Tuller, S. van Dijken, G. S. D. Beach, Magneto-ionic control of interfacial magnetism. *Nat Mater* **14**, 174–181 (2015).

49. G. Huang, Z. Liu, L. Van Der Maaten, K. Q. Weinberger, "Densely Connected Convolutional Networks" in *2017 IEEE Conference on Computer Vision and Pattern Recognition (CVPR)* (IEEE, 2017; https://ieeexplore.ieee.org/document/8099726/)vols. 2017-January, pp. 2261–2269.

50. D. P. Kingma, J. Ba, Adam: A Method for Stochastic Optimization. arXiv:1412.6980 (2014).

51. N. Srivastava, G. Hinton, A. Krizhevsky, I. Sutskever, R. Salakhutdinov, Dropout: a simple way to prevent neural networks from overfitting. *The Journal of Machine Learning Research* **15**, 1929–1958 (2014).

52. H. H. Tan, K. H. Lim, "Vanishing Gradient Mitigation with Deep Learning Neural Network Optimization" in *2019 7th International Conference on Smart Computing & Communications (ICSCC)* (IEEE, 2019; https://ieeexplore.ieee.org/document/8843652/), pp. 1–4.

53. J. Gou, B. Yu, S. J. Maybank, D. Tao, Knowledge Distillation: A Survey. *Int J Comput Vis* **129**, 1789–1819 (2021).

54. G. Hinton, O. Vinyals, J. Dean, Distilling the Knowledge in a Neural Network. arXiv:1503.02531 (2015).

55. J. H. Cho, B. Hariharan, "On the Efficacy of Knowledge Distillation" in *2019 IEEE/CVF International Conference on Computer Vision (ICCV)* (IEEE, 2019; https://ieeexplore.ieee.org/document/9008764/), pp. 4793–4801.

56. CIFAR-10 and CIFAR-100 datasets. https://www.cs.toronto.edu/~kriz/cifar.html.

57. K. He, X. Zhang, S. Ren, J. Sun, "Deep Residual Learning for Image Recognition" in *2016 IEEE Conference on Computer Vision and Pattern Recognition (CVPR)* (IEEE, 2016; http://ieeexplore.ieee.org/document/7780459/)vols. 2016-December, pp. 770–778.